# Developments of an RFQ Cooler SHIRaC: beam transport and nuclearization

**Ramzi Boussaid**[a*], **Gilles ban** [a], **Gilles Quéméner** [a]

[a] *LPC-IN2P3, ENSICAEN, 6 Boul. Maréchal Juin, 14050 Caen, France.*

*E-mail*: boussaidramzii@gmail.com

**ABSTRACT;**

The development of a new RFQ Cooler, named SHIRaC, was carried out to handle and cool typical SPIRAL 2 beams of large emittances (up to 80 $\pi$.mm.mrad) and high currents (up to 1 µA). SHIRaC is a part of SPIRAL 2 facility at GANIL laboratory in France. Its purposes are to enhance as much as possible the beam quality: transverse geometric emittance of less than 3 $\pi$.mm.mrad and a longitudinal energy spread close to 1 eV, and to transmit more than 60 % of ions.
Numerical simulations and experimental studies have shown that the required beam quality can be reached only in the term of the emittance. The energy spread is very far from expected values. It is sensitive to the space charge and buffer gas diffusion and more importantly to the RF field derivative effect. The latter arises at the RFQ exit and increases with the following RF parameters: the frequency and the amplitude of the RF voltage applied to the RFQ electrodes. Studies allowing to enhance the cooled beam quality, mainly the energy spread reduction, are presented and discussed along this paper. They consist in implementing a miniature RFQ at the RFQ exit. Using the development of these studies, it becomes possible to enhance the cooled beam quality and to reach 1 eV of longitudinal energy spread and less than 1.8 pi.mm.mrad of transverse geometric emittance for beam currents going up to 1 µA.
The transport of the cooled beam from the exit of the extraction section towards a HRS has been done with an electrostatic quadrupole triplet. Simulations and first experimental tests showed that more than 95 % of cooled beams can reach the HRS.
Finally, developments related to the nuclearization protection methods aimed to avoid the escape of any nuclear matter from the SHIRaC beamline are studied.

*Ramzi Boussaid, boussaidramzii@gmail.com

# Contents

## Introduction

In the context of the new generation of nuclear facility for the production of rare and exotic ion beams of intensities up to 1µA and emittances up to 80 π.mm.mrad [1], SPIRAL-2 project is currently installed at GANIL laboratory in France [2,3,4]. Such ion beams will subsequently suffer from isobaric contaminations [5].

SPIRAL-2 is devoted to providing beams to DESIR (Desintegration, Excitation and Storage of Ion Radioactive) experiment [6,7]. As DESIR's typical experiment requires the highest purified beam, a HRS (High Resolution Separator) will be installed in front of DESIR [8, 9]. The nominal HRS working requires beams with low beam quality: transverse emittance ≤ 3 π.mm.mrad and longitudinal energy spread of ~1 eV [8]. To do that, the beams should be cooled before reaching the HRS entrance. The Radio-Frequency Quadrupole Cooler (RFQC) is widely known as the universal cooling technique of ion beams [16].

The RFQCs are already used in several different projects [10][11][12][13]. The existing ones can only handle beams with low intensities (~100 nA) and small emittances (~10 πmmmrad). As the SPIRAL-2 typical beams are of higher currents and larger emittances, a new RFQC called SHIRaC (Spiral-2 High Intensity Radiofrequency Cooler) is developed and tested at LPC-Caen laboratory in France. In addition to its new optics system that captures the largest beam emittance, its vacuum system and RF system were also developed [15]. The vacuum system, based on a differential pumping, allows to reduce the buffer gas pressures to less than

0.01 Pa beyond the RFQ chamber. Whereas, with the developed RF system, highest RF parameters of up to 9 MHz and 9 kV of frequency and amplitude, respectively, can be reached without any breakdown limitations. These RF parameters provide the needed RF confinement field to overcome the space charge effects, due to the high beam currents, along the RFQ. This Cooler is devoted to provide the beam quality required by the HRS as well as to transmit more than 60 % of incoming ions toward the HRS.

Simulation [14] and experimental studies [15] of SHIRaC prototype has been investigated and showed pertinent results. The prototype allows to handle and to cool beams of currents going up to 1 µA which has never been achieved before. Using RF voltage parameters of 4.5 MHz in frequency, 4 kV in amplitude (Mathieu parameter q=0.4) and 2.5 Pa in buffer gas pressure, optimum cooling with high transmission has been achieved. The prototype was able to transmit more than 70 % of incoming ions and provide cooled beam quality with less than 2.5 π.mm.mrad of geometric transverse emittance and less than 6.5 eV of longitudinal energy spread. The cooled beam quality is in agreement with the HRS requirements only in the term of emittance. The second beam quality term is much greater than the expected values and its reduction is mandatory for an optimum running of the HRS.

These studies showed that the degrading effects (the space charge, the RF heating and the buffer gas diffusion) have a considerable impact on the cooling process and thereafter on the beam quality [14][15]. Another pertinent degrading effect may occur at the RFQ exit and acts to grow the longitudinal energy spread. A description and analysis of the origin of this effect will be investigated in the first section of this paper.

Study and development of an engineering technique to reduce this pertinent effect and later to avoid the degradation of the cooled beam quality beyond the RFQ end are reported in this paper. Additional special considerations to reach cooled beam quality of 1 eV of longitudinal energy spread and 1 π.mm.mrad of transverse geometric emittance with higher ion transmission are investigated.

In order to transport the cooled beam from the extraction section exit toward the HRS entrance, an electrostatic quadrupole triplet was developed. More details about the simulations studies and experimental tests of the triplet are also presented.

Our purpose to study protection methods against the radioactivity effect, due to the high intensity of SPIRAL 2 typical ion beams, is illustrated.

## 1. Physics motivation: why the DE is so large?

As a recall, SHIRaC is composed of three separate sections: the injection section, the RFQ section and the extraction section [14]. These sections are connected by exchangeable apertures where the buffer gas pressure distributions are obtained by a differential pumping system.

Following the cooling and the transmission through the RFQ section, the cooled beam is extracted then guided with a small acceleration (only few eV) to the extraction plate exit. Once the beam passes through this plate, it is strongly accelerated [14]. The most critical point in the acceleration of the beam is the area between the RFQ exit and the extraction plate exit, where the pressure is extremely high, close to the RFQ pressure [15]. The good beam quality achieved by the cooling could be partially lost due to undesirable collisions of the extracted ions with the buffer gas atoms. The problem worsens in the absence of a confinement field. In addition, the space charge will result in significant degradations as long as the ions' energy is very low.

Comparative simulation results of the beam quality [14], at the RFQ exit and at the extraction section exit, versus the beam currents showing these degradations, are presented in figure 1. Such degradations can be estimated by relying on the relative differences between the beam quality at the RFQ exit and the beam quality at the extraction section exit. Degradations stemming from the buffer gas diffusion correspond to the gap between the curves at low beam current, less than 100 nA. These degradations revolve around the emittance and more importantly around the energy spread. They are worth 0.2 π.mm.mrad and 2.5 eV, respectively.

The gap widens progressively with the beam currents while the space charge effect increases with them. The arising of this phenomenon can be observed in the quick degradation of the emittance with the beam currents (figure 1-left). The energy spread is in the same way affected as it increases (figure 1-right). Its degrading behavior is also observed experimentally as seen on figure 1-right. We also note a good agreement between the simulation and experimental results.

The occurring degradation at low beam current (less than 100 nA), where the space charge effect can be neglected, is due to buffer gas diffusion and is at 10 % for the emittance and at more than 300 % for the energy spread. The energy spread degradation is too large to be stemming only from the buffer gas diffusion because, on one hand, the obtained energy spread is much greater than those provided with the existing cooler, notwithstanding that the buffer gas pressure is at the same level and that the vacuum system is not as powerful as the SHIRaC's [17] and, on the other hand, the gas diffusion had a small effect on the beam energy that did not exceed 10 %.

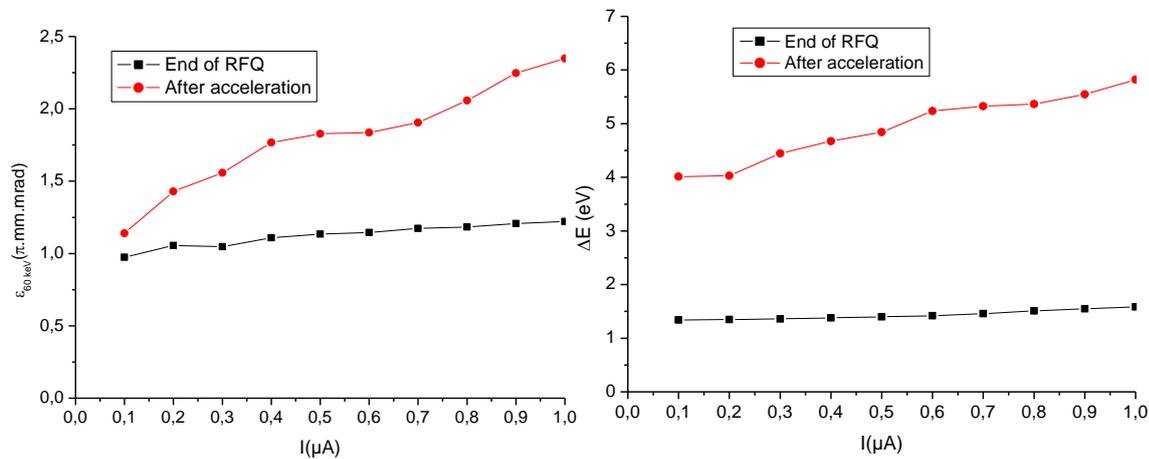

Figure 1: space charge effect on the beam quality: simulation of the beam current effect on the beam emittance (right) and longitudinal energy spread at the RFQ section exit and at the exit of the RFQ Cooler line (left) [14].

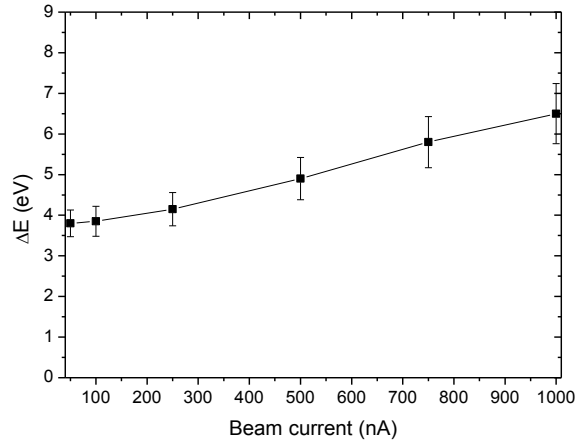

Figure 2: space charge effect on the longitudinal spread energy: experimental and simulation results of the beam current effect on the longitudinal energy spread at the exit of the RFQ Cooler line [14][15].

The main difference between previous RFQCs and SHIRaC lies in the RF voltage parameters used, which are less than 1 kV and 1 MHz of RF voltage amplitude and frequency, respectively, in the former [24], compared to more than 2.5 kV and 4.5 MHz in the latter. This leads us to study the RF parameters' effect on the energy spread. In figure 3, we present the variation of ΔE as a function of the RF voltage amplitude for beam current of 1 μA. Contrary to expected behavior of cooled beam quality with the $V_{RF}$, an increase of the energy spread occurs. This increase is not explained by the effect of the RF voltage amplitude on the cooling (i.e., on the ΔE). Rather, the RF field derivative effect at the RFQ exit is used to explain it. The derivative of the transverse RF field at the RFQ exit gives rise to a longitudinal RF field which acts on the longitudinal velocities distribution of the cooled ions via its random characteristic. Thereafter, it results in a degradation of this distribution and, consequently, in growth of the energy spread.

The derivative of transverse RF field at the RFQ exit can possesse a transverse component. This component has the same confinement effect as the RF field into the RFQ, but with smaller power. This explains that the small degradation of the emittance with low beam current, which occurred after the RFQ end (figure 1), is due to the buffer gas diffusion.

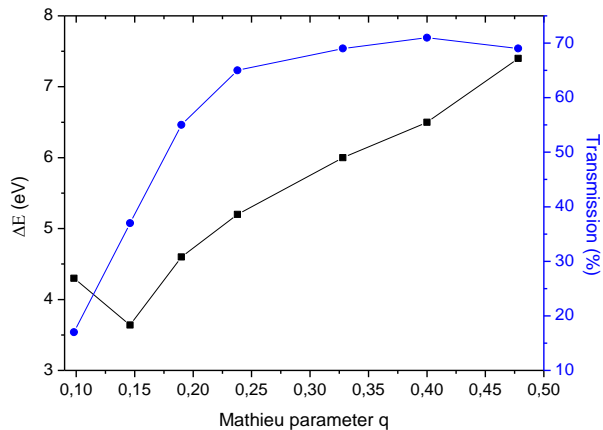

Figure 3: confinement RF voltage amplitude effect on the longitudinal energy spread: variation of the longitudinal energy spread with the RF voltage amplitude [15].

To better understand about the derivative effect of the RF field, the variation of the longitudinal energy spread with the RF voltage frequency was experimentally studied, as shown in figure 4. At the RFQ exit, while the cooled beams energy is very slow, they are under the influence of collective effects, i.e., the ions are into the RF derivative field. This gives rise to ion fluctuations and subsequently to beam quality growth. This phenomenon is very clear on figure 4 where the energy spread increases with the RF voltage frequency. Such increase is due to the growth of the ions fluctuation rate. To conclude, the derivative effect arose and grew with the RF voltage parameters.

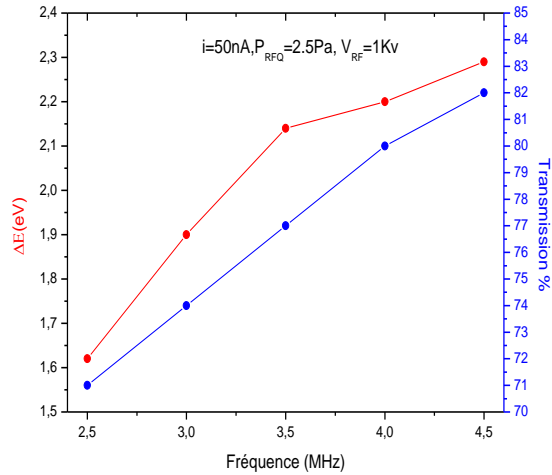

**Figure** 4: collective effect on the longitudinal spread energy: variation of the DE versus the RF voltage frequency.

## 2. Extraction section development

As cited above, the space charge and the buffer gas diffusion at the RFQ exit have a contribution in the energy spread growth. However, the RF field derivative effect has the greatest contribution in this degradation. In order to avoid this degradation, an optical device providing both the transport of cooled ions from one vacuum stage to another, with only minor disturbances, and a confinement field, should be implemented at the RFQ exit. One of the options is to set up a miniature RFQ (µRFQ).

### 2.1. Miniature RFQ Cooler (µRFQ)

The µRFQ device technique was used in several projects [20] [18, 19] as a method allowing to just reduce the buffer gas diffusion. In the case of SHIRaC, this device should furthermore be able to:
- Overcome the space charge effects which occur at the RFQ exit.
- Reduce the RF field derivative effect: It acts as a cut-off of the RF voltage derivative propagation.
- Guide the cooled ion beam to a region where their energy is of a few tens eV, so then they can resist to any degrading effects.

Simulation studies of this µRFQ have been done, related to the optimization of its design as well as its position relative to the RFQ end. These simulations intend to study, likewise, the cooled beam quality dependencies on the RF voltage amplitude $V_{RF}$, the RFQ buffer gas pressure $P_{RFQ}$ and the space charge. They are based on the SIMION 3D V8.0 program and the hard disc model as presented in reference [25][26]. $_{133}Cs^+$ ion beams of 60 keV of energy and

80 π.mm.mrad of geometric transverse emittance were used for this. Preliminary simulations showed that the optimized µRFQ dimensions are 40 mm of length and 2 mm of inner radius. In order to keep the lowest buffer gas diffusion whilst avoiding any breakdown between the electrodes of the RFQ and those of the µRFQ, the µRFQ electrodes are half circular. With this method, the pressure distribution beyond the RFQ chamber is reduced of about 10 %.

In order to avoid the RF field derivative effect and any acceleration or deceleration phenomenon of ions in the interface between the RFQ and the µRFQ, the latter must run with the same frequency as the RFQ's. For ions stability reasons along the µRFQ, low RF voltage amplitudes supplying their electrodes should be used. Thus, the µRFQ must operate with RF amplitude of 441 V for a frequency of 4.5 MHz, hence a Mathieu parameter q=0.4. In order to allow the operator to make use of the same RF system for both the main RFQ and the µRFQ, a classic voltage divider should be installed.

In order to adequately extract the cooled beam from the µRFQ while avoiding any degrading effect on the obtained cooled beam quality, double extraction electrodes of 3 mm of inner radius are installed at the µRFQ exit (figure 5). The first electrode is 1 mm distant from the µRFQ end and 2 mm from the second electrode.

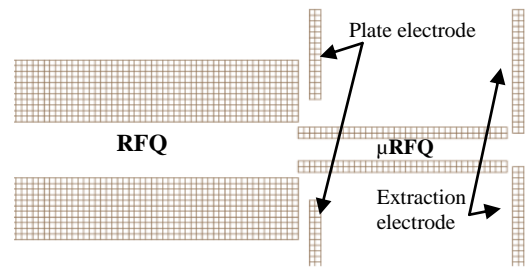

Figure 5: relative positioning of the µRFQ from the RFQ and layout of the new extraction section.

## 2.2. Longitudinal energy spread improvement

Using the new design of the extraction section, simulation results of the space charge and RF field derivative effects on the cooling process are illustrated in this section. The longitudinal energy spread ΔE were investigated as a function of beam currents and RF voltage amplitudes.

The below figure illustrates simulated results of the longitudinal energy spread as a function of the RF voltage amplitude (i.e., the Mathieu parameter q), applied to the main RFQ electrodes, while keeping constant the RF voltage amplitude (satisfying q= 0.4) applied to the µRFQ. The RF field derivative effect is clearly deleted because the longitudinal energy spread doesn't increase with the RF voltage amplitude, as seen in figure 2, but rather decreases until reaching a plateau for q between 0.3 and 0.5. The plateau mirrors the obtained optimum cooling. This behavior of ΔE reveals very clearly, on one hand, the expected RF voltage amplitude effect on the confinement power and thereafter on the beam cooling throughout RFQ section, and shows the cut-off of the field derivative effect by the µRFQ on the other hand. The increase in the ΔE for q > 0.5 is explained by the occurrence of the RF heating effect on the cooling.

Besides the removing of the derivative effect, we clearly note that both the space charge and the buffer gas diffusion effect are reduced in view of lower energy spread values.

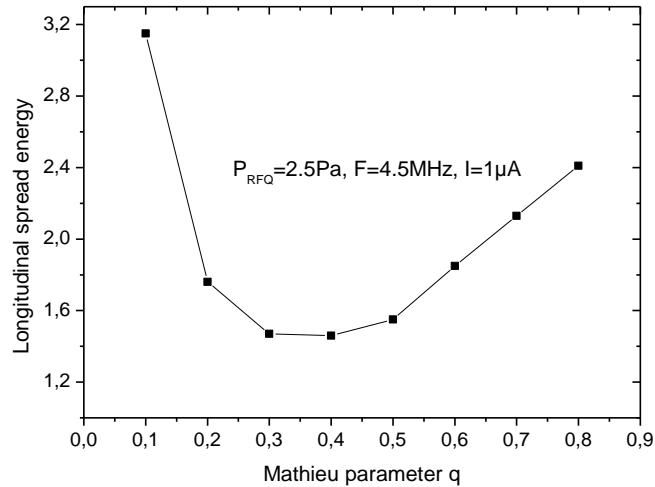

Figure 4: RF voltage effect on the longitudinal energy spread for 1µA ion beam: variation of the longitudinal energy spread as a function of the Mathieu parameter q.

The µRFQ contribution to reduce the space charge effect on the longitudinal energy spread should be noticeable in view of the presence of a confinement field to overcome this effect as long as the cooled ions are slow at the RFQ exit. This is very clear, on one hand, in figure 5-Left, as the ΔE increases little by little from 1.1 eV to 1.4 eV with beam currents going up to 1 µA. It is visibly manifested in the reduction of the variation gap of the ΔE, which was more than 2.5 eV without the µRFQ [14] versus only 0.3 eV with the present developments, for currents going from 50 nA to 1 µA. On the other hand, it is also clearly involved as the energy spread values are very smaller than those without the µRFQ, such as the ΔE that is reduced from 6.45 eV to 1.48 eV for 1 µA.

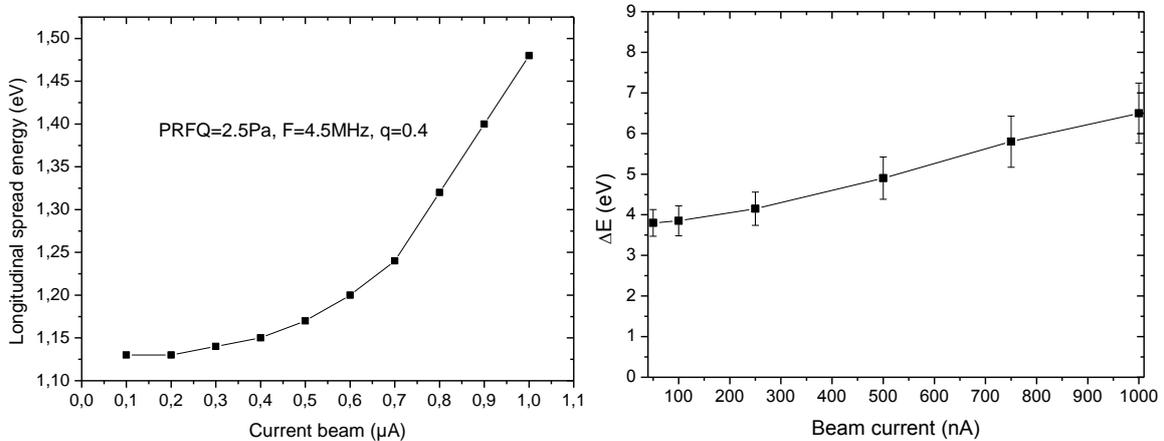

Figure 5: Space charge effect on the longitudinal energy spread for q=0.4.

### 2.3. Cooled beam quality improvement

As showed above, the µRFQ contribution is very efficient to reduce as low as possible the degrading effects of the cooled beam quality. A supplementary cooling of ions when crossing traversing the µRFQ can take place. The cooling can be improved in increasing the buffer gas pressure while maintaining low buffer gas diffusion. Thus, any increase of RFQ pressure results in enhancing the cooled beam quality. The below figure illustrates this phenomenon as the minimum longitudinal energy spread, around 1.01 eV, is obtained with 4 Pa of pressure, instead of 2.5 Pa without the µRFQ. As is expected, ΔE is improving with the buffer gas pressure until reaching its optimum value at 4 Pa where the cooling is optimum. The increase in the ΔE at the high pressures, beyond 4.5 Pa, appeared to come from the occurrence of the buffer gas diffusion effect at the µRFQ exit.

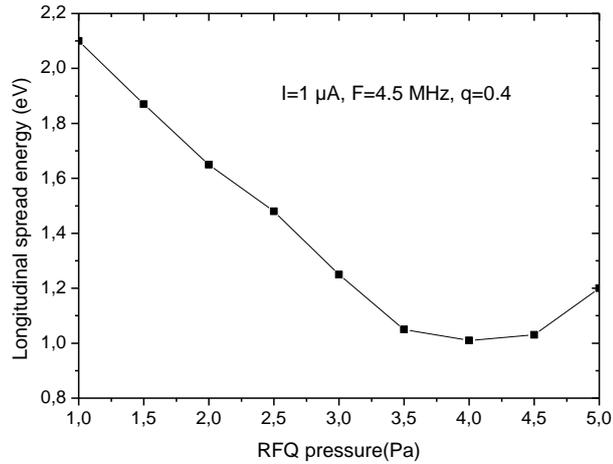

Figure 6: RFQ pressure effect on the longitudinal spread energy

The cooling improvement with the buffer gas pressure should automatically result in an emittance reduction. The geometric transverse emittance, defined as the area of an ellipse containing more than 90 % of the action volume of cooled ions distribution, is determined with 4 Pa of buffer gas pressure. Its reduction is very clear where it decreases from 2.5 π.mm.mrad without the µRFQ [15] to around 1.8 π.mm.mrad respectively, with 2.5 and 4 Pa. It is also important to note that the simulations showed the transmissions remain unchangeable in implementing the µRFQ and exceed 70 % for beam current going up to 1 µA.

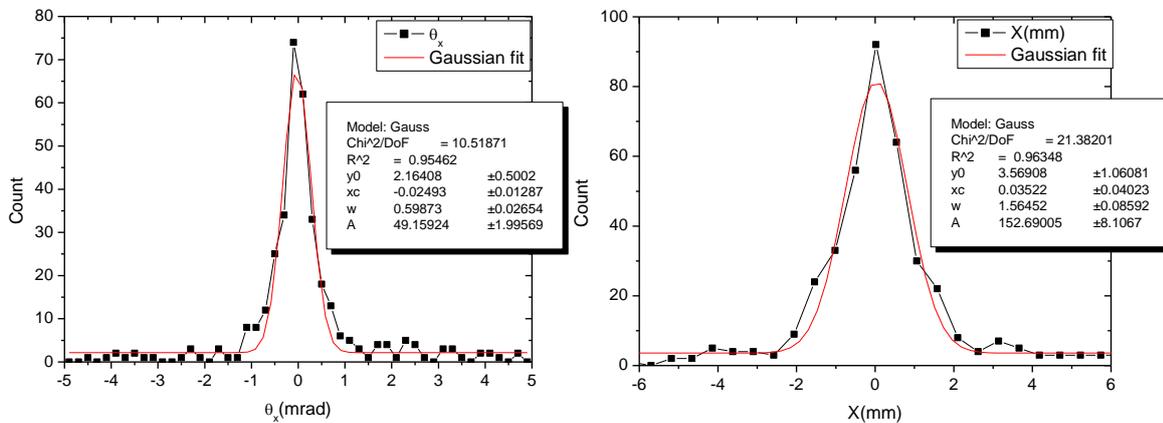

Figure 7: characteristics of 1µA cooled ion beam: distributions of the elevation angle $\theta_X$(left) and the distribution of the radial position X(right).

## 3. Coupling RFQ–HRS: quadrupole triplet

The cooled beams outgoing from SHIRaC are accelerated toward the HRS entrance where they should pass through a rectangular slit which dimension is 1×4 mm². Without any ion optical device, the ions losses are significant and less than 20 % of them can access the HRS. In order to avoid these losses, the device should be installed between the extraction section exit and the HRS entrance and must allow focusing the beam at that slit. It also must be run with only a few kilovolts of voltage DC (less than 5 kV) for energy beams going up to 60 keV. The suitable solution is a multiple of electrostatic quadrupole which can provide focus that is stronger than the Einzel lens [21].

Because the tuning of a quadrupole triplet is much easier than other quadrupole multiplet and because, at the same time, a triplet remains flexible enough to satisfy various requirements in the first order focusing properties, a quadrupole triplet have been adopted in the present work. The design of the triplet and the optimized dimensions of their quadrupoles are done by Simion 3D V8.1. The optimum design is shown in figure 8 and their dimensions are illustrated in table 1.

Simulations have shown important aberrations of the beam through the triplet, due to the edge effects of their electrodes. In order to reduce these aberrations and to protect the electrodes from beam, collimators were installed on either side of each quadrupole [ref].

Voltages are applied to the triplet in such a way that, in both x–z and y–z planes, converging and diverging lenses alternate. The y–z plane has been taken to be the DCD (diverging–converging–diverging) plane and x–z plane to be the CDC (converging–diverging–converging) plane. The optimum voltages for a beam of 5 keV are of 130 V, 130 V and 220 V respectively for the first, second and third quadrupoles. However, for a beam of 60 keV they are of 2.3 kV, 2.3 kV and 4 kV respectively.

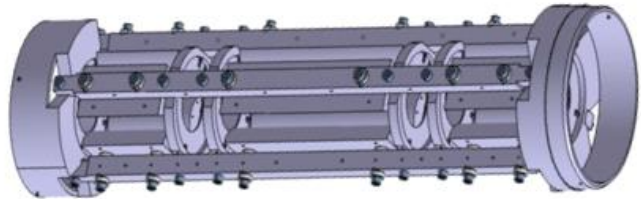

Figure 8: Schematic design of electrostatic quadrupole triplet

| Compounds | Dimensions (mm) |
|---|---|
| 1$^{st}$ quadrupole (EQ 1) | 80 |
| Drift: EQ 1- EQ 2 | 60 |
| 2$^{nd}$ quadrupole (EQ 2) | 160 |
| Drift: EQ2-EQ 3 | 60 |
| 3$^{rd}$ quadrupole (EQ 3) | 80 |
| Drift: EQ 3 - HRS entrance | 230 |
| Total length | 440 |
| Inner radius | 15 |
| Length of the collimator | 2 |
| Inner radius of the collimator | 15 |

**Table 1**: specification of the electrostatic quadrupole triplet.

In the below figure, we have presented the measurements of the transverse cooled beam profiles at the DCD and CDC planes for various beam currents. The space charge effect on the transverse size of the beams is clear while there are widening of these profiles.

To quantitatively study this phenomenon, the FWHM ($\sigma_x$ or $\sigma_y$) of these profiles were determined, as presented in table 2. For beam currents going up to 1μA, the triple of these FWHM (3.$\sigma_x$ or 3.$\sigma_y$) does not exceed the dimension of the slit (3.$\sigma_x$ < 1 mm and 3.$\sigma_y$<4 mm). Therefore, more than 95% of cooled beams can pass through the slit toward the HRS

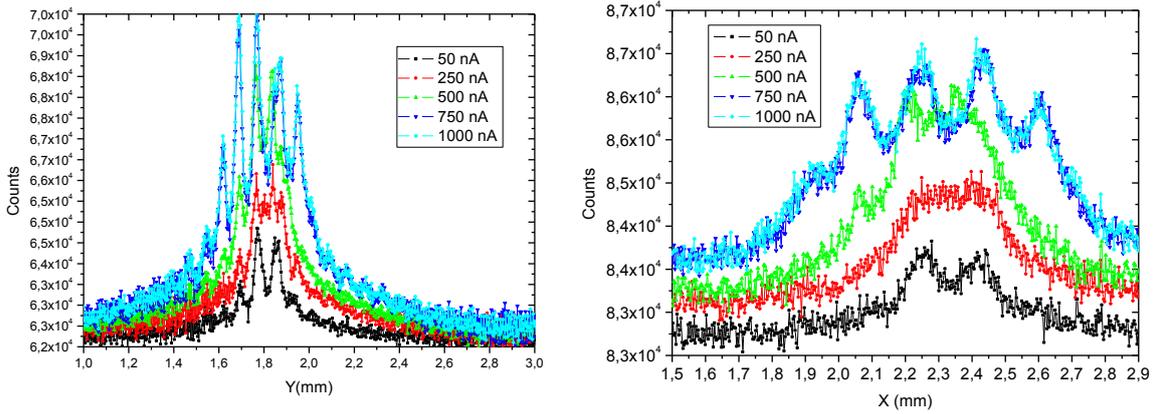

Figure 9: Transverse profiles of the cooled beams measured at the exit of the HRS slit for various beam currents: profile at the CDC plan (left) and its counterpart at the DCD plan (right).

| Intensity (nA) | $\sigma_x$ (mm) | $\sigma_y$ (mm) |
|---|---|---|
| 50 | 0.231 | 0.347 |
| 250 | 0.246 | 0.350 |
| 500 | 0.242 | 0.362 |
| 750 | 0.308 | 0.579 |
| 1000 | 0.374 | 0.586 |

Table 2: The FWHM (Full half high Medium) of Gaussian fits of the cooled beam profiles for various beam currents.

## 4. Nuclearization : confinement of nuclear matter

As mentioned above, SHIRaC will be implemented within the SPIRAL 2 line and will receive their beams. Because these beams will be of highest currents, going up to 1 µA [1], a high radioactivity environment into the experimental hall of SHIRaC would be very dangerous. To avoid this problem, specific considerations to confine the nuclear matter (radioactive chemical compounds) should be studied and developed. This requires, aside the buffer gas recycle, the usage of appropriate safety processes.

The confinement of the nuclear matter consists in avoiding the escape of any nuclear matter during the experimental test, mostly at the maintenance intervention. As the used vacuum system [14] ensures the statistic confinement of the nuclear matter, the dynamic confinement demands specific turbo molecular pumps with double valves and axial collar for the clamp (figure 9). The first valve is fixed to the module and the second one to the beam-line. The sweep of gussets between the double valves is performed before the disconnection of the module. The gas recovered following this sweep is analyzed and, if necessary, stored for radioactive decay.

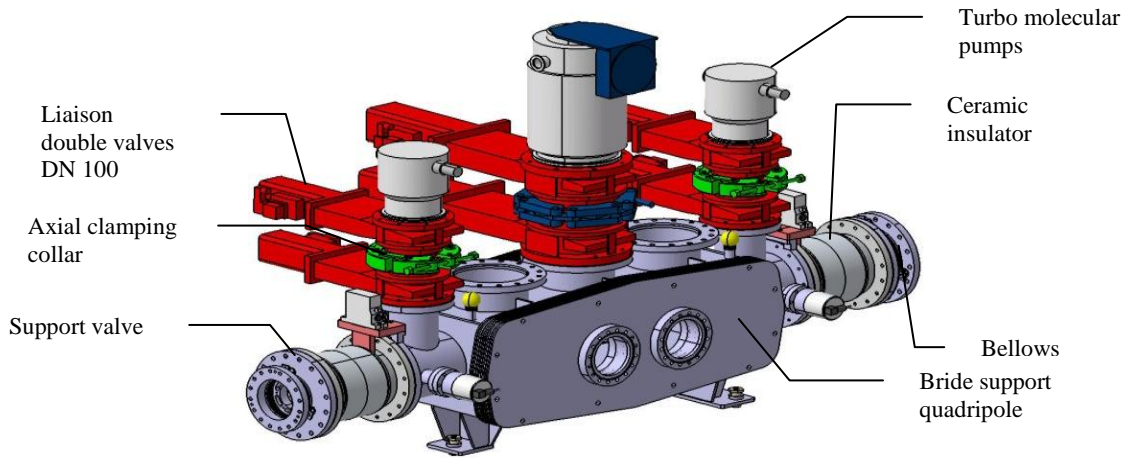

Figure 9: new design of the RFQ chamber of SHIRaC: dynamic confinement method of the nuclear matter by turbo-molecular pumps with double valves and axial collar for clamp.

To maintain the RFQ module, withdraw or create elements under vinyl, the nuclear matter should be confined into a volume. This volume can occur via a confinement vinyl. Before opening the RFQ chamber, we should setup a waterproof sleeve which allows to avoid the escape of the nuclear matter. The steps of this process are done as follows (Figure 1): First, we set up a confinement vinyl. Then, in order to create two distinct volumes, we withdraw the RFQ support and, finally, we separate the two subsets. During the implementation phase of the RFQ, we should, foremost, setup the waterproof sleeve and then withdraw it before mounting the RFQ.

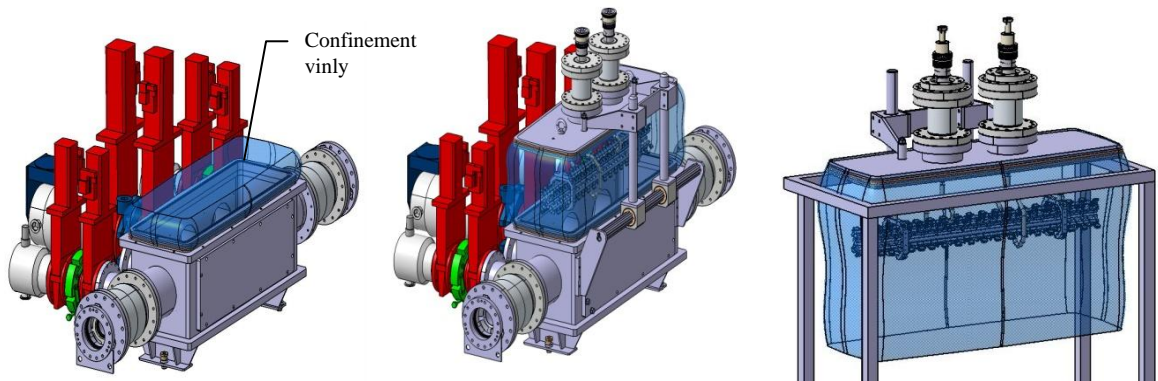

Figure 10: steps of the RFQ withdraw: Establishment of confinement vinyl (left), withdraw and creation of two volumes (middle) and separation of two subsets (right).

## Conclusion and perspectives

The ingenious solution of the µRFQ has showed a high efficiency in enhancing the cooled beam quality and in reducing both the emittance and the longitudinal energy spread. The cooled beam quality reached the expected values of SHIRaC requirements and thereafter an isobaric purification with the HRS will be possible.

The proposed solutions for the nuclearization protection and the µRFQ presented in this paper are under realization.

Owing these studies, SHIRaC prototype will provide cooled beam quality never reached so far, and beam currents never handled before. However, the cooling of beam currents exceeding 1 µA will need a new design of RFQ Cooler and even the µRFQ will not be able to overcome the degrading effects. Special considerations of the extraction section should be studied for the next generation of RFQ Cooler for the EURISOL project [22][23] where the beam current will reach 10 µA.

**Acknowledge:**

We would like to thank Professor G.Ban for continuous encouragement and support. The teams of electronic, vacuum system and mechanical design at L.P.C Caen (France) are gratefully acknowledged for their kind assistance in the development of the project. We also thank Mouna yahyaoui for her English corrections.

# References

[1] S.Gales., SPIRAL 2 at GANIL: Next generation of ISOL Facility for Intense Secondary Radioactive Ion Beams, Nucl. Phys A 834 (2010) 717c–723c

[2] Letter of intent for SPIRAL 2.

[3] M.Lewitowicz., The SPIRAL 2 Project, Nucl. Phys. A 805 (2008) 519-525c, GANIL

[4] M.Lewitowicz, STATUS OF THE SPIRAL2 PROJECT, ACTA Physica Polonica B, Vol 42(2011)

[5] E.M. Ramirez et al,. The ion circus: A novel circular Paul trap to resolve isobaric contamination, Nucl. Instrum. Meth. B 266 (2008) 4460.

[6] The DESIR facility, Letter of intents for SPIRAL 2 (2006)

[7] B.Blanc et al., DESIR: The SPIRAL2 Low-Energy Beam Facility, Technical Proposal for SPIRAL2 Instrumentation, 2008, pp. 1-102

[8] D. Toprek et al., DESIR high resolution separator at GANIL (France), Nucl. Tech. Rad. 27 (2012) 346.

[9] T.Kurtukian-Nieto et al., SPIRAL2/DESIR high resolution mass separator, Nucl. Instr. And Meth. In Phys. B 317 (2013) 284–289

[10] G. Bollen et al, Penning trap mass measurements on rare isotopes status and new developments, J. Phys. B 36 (5) (2003) 941-951.

[11] M. Mukherjee et al, ISOLTRAP: An on-line Penning trap for mass spectrometry on short-lived nuclides, the european physics journal. A 35, 1–29 (2008)

[12] I.Podadera, New developments on preparation of cooled and bunched radioactive ion beams at ISOL-Facilities: The ISCOOL project and the rotating wall cooling, PhD thesis, Universitat Politècnica de Catalunya, Barcelona Spain (2006).

[13] J. Dilling et al., Mass measurements on highly charged radioactive ions, a new approach to high precision with TITAN., Int. J. Mass. Spectrom. 251, 198-203 (2006).

[14] R.Boussaid et al., Simulations of high intensity ion beam RFQ Cooler for DESIR/SPIRAL2 : SHIRaC, J.INST 9 P07009. Doi:10.1088/1748-0221/9/07/P07009, (2014). http://dx.doi.org/10.1088/1748-0221/9/07/P07009

[15] R.Boussaid et al., Experimental study of a high intensity radiofrequency cooler, accepted by PRSTAB journal: http://arxiv.org/abs/1403.1148.

[16] R.B.Moore et al, Improving isotope separator performance by beam cooling, Nucl. Instr. and Meth. in physics research B 204 (2003) 557-562.

[17] I.Podadera, New developments on preparation of cooled and bunched radioactive ion beams at ISOL-Facilities: The ISCOOL project and the rotating wall cooling, PhD thesis, Universitat Politècnica de Catalunya, Barcelona Spain (2006).

[18] F.Herfurth et al., Nucl. Instr. And Meth. In Phys. A 469 (2001) 254-275.

[19] S. Schwarz et al,. Nucl. Instr. And Meth. In Phys. B 204 (2003) 474–477.

[20] T. Kim, Buffer gas cooling of ions in a radiofrequency quadrupole ion guide: a study of the cooling process and cooled beam properties, PhD. thesis, McGill University, 1997.

[21] Q.Zhao et al., MSU, East Lansing, MI 48842, U.S.A. Proceedings of 2005 Particle Accelerator Conference, Knoxville, Tennessee.

[22] M.Lewitowicz, The spiral 2 project, Nuclear physics A 805 (2008) 519-525c, GANIL

[23] J.Vervier, Status report of EURISOL project, Nuclear Instruments and Methods in Physics Research Section B 204(759– 764) 2003.

[24] O. Gianfrancesco et al, A radiofrequency quadrupole cooler for high-intensity beams, Nucl. Instr. and Meth. in Phys. Res. B 266 (2008) 4483–4487

[25] D.Manura. Additional Notes on the SIMION HS1 Collision Model. 2007, Scientific Intrument Services.

[26] D.A.Dah,Simion 3D V8.0 User Manual, Idaho National Engineering Laboratory (2000).